\documentclass[aps,prd,twocolumn,superscriptaddress,bibnotes,longbibliography,preprintnumbers,floatfix,10pt]{revtex4-2}

\usepackage{graphicx}
\usepackage{xcolor}
\usepackage{amsmath,amsfonts,amsthm,amssymb}
\usepackage[colorlinks]{hyperref}
\usepackage{mathtools}
\usepackage{bm}
\usepackage{multirow}
\makeatletter

\newcommand{\orcid}[1]{\href{https://orcid.org/#1}{\includegraphics[width=10pt]{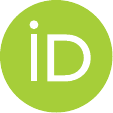}}}

\hypersetup{
    pdfnewwindow=true,      
    colorlinks=true,       
    linkcolor=violet,          
    citecolor=violet,        
    filecolor=violet,      
    urlcolor=violet        
}

\begin{document}

\title{Enhancing DUNE's solar neutrino capabilities with neutral-current detection}

\author{Stephan A. Meighen-Berger \orcid{0000-0001-6579-2000}\,}
\email{stephan.meighenberger@unimelb.edu.au}
\affiliation{School of Physics, The University of Melbourne, Victoria 3010, Australia \href{https://ror.org/01ej9dk98}{[ROR]}}
\affiliation{Center for Cosmology and AstroParticle Physics (CCAPP), Ohio State University, 
Columbus, OH 43210, USA \href{https://ror.org/00rs6vg23}{[ROR]}}

\author{Jayden L. Newstead \orcid{0000-0002-8704-3550}\,}
\email{jnewstead@unimelb.edu.au}
\affiliation{School of Physics, The University of Melbourne, Victoria 3010, Australia \href{https://ror.org/01ej9dk98}{[ROR]}}
\affiliation{ARC Centre of Excellence for Dark Matter Particle Physics, School of Physics, The University of Melbourne, Victoria 3010, Australia \href{https://ror.org/00rs6vg23}{[ROR]}}

\author{\\John F. Beacom \orcid{0000-0002-0005-2631}\,}
\email{beacom.7@osu.edu}
\affiliation{Center for Cosmology and AstroParticle Physics (CCAPP), Ohio State University, Columbus, OH 43210, USA \href{https://ror.org/00rs6vg23}{[ROR]}}
\affiliation{Department of Physics, Ohio State University, Columbus, Ohio 43210, USA \href{https://ror.org/00rs6vg23}{[ROR]}}
\affiliation{Department of Astronomy, Ohio State University, Columbus, Ohio 43210, USA \href{https://ror.org/00rs6vg23}{[ROR]}}

\author{Nicole F. Bell \orcid{0000-0002-5805-9828}\,}
\email{n.bell@unimelb.edu.au}
\affiliation{School of Physics, The University of Melbourne, Victoria 3010, Australia \href{https://ror.org/01ej9dk98}{[ROR]}}
\affiliation{ARC Centre of Excellence for Dark Matter Particle Physics, School of Physics, The University of Melbourne, Victoria 3010, Australia \href{https://ror.org/00rs6vg23}{[ROR]}}

\author{Matthew J. Dolan \orcid{0000-0003-3420-8718}\,}
\email{matthew.dolan@unimelb.edu.au}
\affiliation{School of Physics, The University of Melbourne, Victoria 3010, Australia \href{https://ror.org/01ej9dk98}{[ROR]}}
\affiliation{ARC Centre of Excellence for Dark Matter Particle Physics, School of Physics, The University of Melbourne, Victoria 3010, Australia \href{https://ror.org/00rs6vg23}{[ROR]}}

\date{\today}

\begin{abstract}
We show that the Deep Underground Neutrino Experiment (DUNE) has the potential to make a precise measurement of the total active flux of $^8$B solar neutrinos via neutral-current (NC) interactions with argon.  This would complement proposed precise measurements of solar-neutrino fluxes in DUNE via charged-current (CC) interactions with argon and mixed CC/NC interactions with electrons.  Together, these would enable DUNE to make a SNO-like comparison of rates and thus to make the most precise measurements of $\sin^2\theta_{12}$ and $\Delta m^2_{21}$ using solar neutrinos. Realizing this potential requires dedicated but realistic efforts to improve DUNE's low-energy capabilities and separately to reduce neutrino-argon cross section uncertainties.  Comparison of mixing-parameter results obtained using solar neutrinos in DUNE and reactor antineutrinos in JUNO (Jiangmen Underground Neutrino Observatory) would allow unprecedented tests of new physics.
\end{abstract}

\maketitle


\section{Introduction}

An era of radically improved precision is beginning in neutrino physics.  At present, the neutrino-mixing parameters are known to $\sim$3--10\% precision~\cite{PDG2024}.  In 2024, the Jiangmen Underground Neutrino Observatory (JUNO) will begin its measurements of reactor antineutrinos, achieving sub-percent level precision for $\Delta m^2_{21}$, $\sin^2 \theta_{12}$, and $\Delta m^2_{31}$~\cite{JUNO:2015zny, JUNO:2022mxj}.  In 2027, the Hyper-Kamiokande experiment will begin several measurements, improving the present $\Delta m^2_{32}$ and $\sin^2\theta_{23}$ uncertainties by a factor of $\sim$3~\cite{Fukasawa:2016yue, Hyper-Kamiokande:2018ofw}.  Then, beginning about 2030, the long-baseline program at the Deep Underground Neutrino Experiment (DUNE) will provide even greater precision~\cite{DUNE:2020jqi, SajjadAthar:2021prg}.

With these huge leaps in precision come huge opportunities to test for new physics in the neutrino sector, i.e., anything beyond simple three-flavor mixing.  The “solar” mixing parameters, $\sin^2 \theta_{12}$ and $\Delta m^2_{21}$, can be measured in two independent ways: with solar neutrinos or reactor antineutrinos.  With the former, we have a long baseline, large matter effects, and large magnetic fields.  With the latter, we have short baselines, and effectively zero matter density and magnetic fields.  To match the opportunity of JUNO's precision reactor neutrino measurements, significantly improved solar neutrino measurements are needed.  With those, we could powerfully test a variety of new-physics scenarios for solar neutrinos~\cite{Barbier:2004ez, Friedland:2004pp, Palazzo:2011rj, Barenboim:2023krl, deGouvea:2023jxn, Ansarifard:2024zxm, Wu:2023twu, Denton:2024upc}.

\begin{figure}[b]
\centering
\includegraphics[width=0.9\columnwidth]{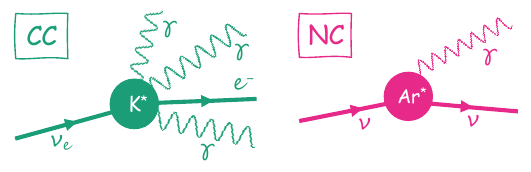}
\caption{Charged-current (left) and neutral-current (right) detection channels in DUNE.}
\label{fig:event_topology}
\end{figure}

Capozzi et al.~\cite{Capozzi:2018dat}, building on prior work on solar-neutrino detection in liquid argon~\cite{Raghavan:1986fg, Bahcall:1986ry, Arneodo:2000fa, Franco:2015pha, Ioannisian:2017dkx}, showed that DUNE has the potential to make the world's most precise measurements of high-energy ($E_\nu > 5$ MeV) solar neutrinos.  In their proposal, $\Delta m^2_{21}$ would be measured through the day-night effect, which is independent of the solar $^8$B flux and the uncertainties on the neutrino-argon cross section, while $\sin^2\theta_{12}$ would be determined by comparing the rates of two channels. The first is charged-current (CC) interactions with argon ($\nu_e + \, ^{40}{\rm Ar} \rightarrow e^- + \, ^{40}{\rm K}^*$, where * indicates a nuclear excited state).  The second is mixed CC/NC elastic scattering (ES) interactions with electrons ($\nu + e^- \rightarrow \nu + e^-$).  This approach breaks the degeneracy between the mixing angle and the $^8$B flux because the high-energy rates depend on the fluxes as:
\begin{eqnarray}
    \mathcal{R}^\mathrm{CC} & \propto & \phi_{\nu_e} \label{eq:CC} \\
    \mathcal{R}^\mathrm{ES} & \propto & \phi_{\nu_e} + \frac{1}{6} \, (\phi_{\nu_\mu} + \phi_{\nu_\tau}), \label{eq:ES}
\end{eqnarray}
where the $\phi_\alpha$ terms indicates the fluxes of the various flavors after neutrino mixing and $\phi_{\rm tot}$ is the total active flux.  Here, Eq.~(\ref{eq:CC}) depends on $\phi_{\rm tot} \, \sin^2\theta_{12}$, while Eq.~(\ref{eq:ES}) depends on $\phi_{\rm tot} \, (\sin^2\theta_{12} + \frac{1}{6} \cos^2\theta_{12})$.  This is not an ideal comparison, due to the $\simeq$1/6 factor, but this can be partially overcome with the expected large statistics.

\begin{figure*}[t]
\centering
\includegraphics[width=\textwidth]{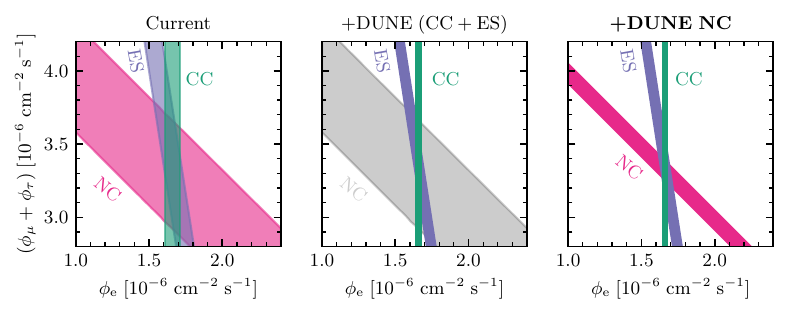}
\caption{The $\nu_e$ and $\nu_{\mu}+\nu_\tau$ content of the solar flux (statistical uncertainties only). {\it Left}: Current measurements from SNO (CC and NC)~\cite{SNO:2011hxd} and Super-Kamiokande (ES)~\cite{Super-Kamiokande:2016yck, Super-Kamiokande:2023jbt}. {\it Middle}: Predictions for DUNE CC+ES from Ref.~\cite{Capozzi:2018dat}. {\it Right}: Adding the DUNE NC measurement proposed here.  \textit{Measurement of the NC rate in DUNE would have a major impact.}}
\label{fig:normalization}
\end{figure*}

In this paper, we show that DUNE has the potential to do even better by also using the pure NC channel ($\nu + \, ^{40}{\rm Ar} \rightarrow \nu + \, ^{40}{\rm Ar}^*$), where the rate is
\begin{eqnarray}
    \mathcal{R}^\mathrm{NC} & \propto & \phi_{\nu_e} + \phi_{\nu_\mu} + \phi_{\nu_\tau} \equiv \phi_{\rm tot}, \label{eq:NC}
\end{eqnarray}
independent of the details of active-flavor mixing.  Importantly, the ratio $\mathcal{R}^\mathrm{CC}/\mathcal{R}^\mathrm{NC}$ directly determines $\sin^2\theta_{12}$, providing a conceptually cleaner way to measure the mixing angle.  Moreover, by providing a third type of measurement, it provides a cross-check and improves the overall precision. 

Figure~\ref{fig:event_topology} sketches how CC and NC events can be distinguished.  Since earlier work, where the NC channel was only briefly discussed for solar neutrinos~\cite{Raghavan:1986fg, Capozzi:2018dat}, there have been two key developments that make it seem much more promising.  First, the experimental work of Refs.~\cite{PhysRevC.73.054306, Gayer:2019eed, Tornow:2022kmo}, which have measured the most important nuclear excitations in argon, significantly increasing the precision of the NC cross section; this shows that further improvements are feasible.  Second, increased optimism about the ability for liquid-argon detectors to register nuclear de-excitations through the emission of MeV-range gamma rays~\cite{ArgoNeuT:2018tvi, Castiglioni:2020tsu, Gardiner:2020ulp}.

Figure~\ref{fig:normalization} (details given in Sec.~\ref{sec:DUNE_prospects}) shows the sensitivity to the solar neutrino fluxes in three stages: with current data, including the precise ES measurement by Super-Kamiokande~\cite{Super-Kamiokande:2016yck} and the pioneering CC and NC measurements in heavy water in the Sudbury Neutrino Observatory (SNO)~\cite{SNO:2011hxd}; with the ``DUNE Solar" capabilities proposed in Ref.~\cite{Capozzi:2018dat}; and with the advances proposed here via using the NC channel too. This final step would enable DUNE to make a direct and powerful SNO-like CC/NC comparison.  \textit{Our primary goal in this paper is to demonstrate the physics potential of NC measurements in DUNE while pointing out key experimental challenges.}

In Section~\ref{sec:nuclear}, we review the physics of neutrino-argon interactions and explain the new developments.  Then, in Section~\ref{sec:DUNE_prospects}, we present our main calculations for NC interactions in DUNE, the implications for solar neutrinos, and the required next steps.  In Section~\ref{sec:conclusion}, we conclude.  In Appendices, we provide further information.


\section{Neutrino-Argon NC Interactions}
\label{sec:nuclear}

\begin{figure}[t]
\centering
\includegraphics[width=0.6\columnwidth]{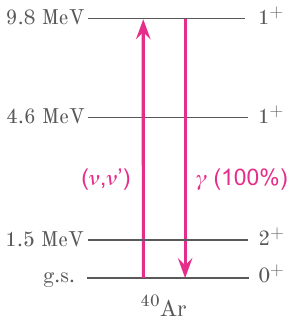}
\caption{Greatly simplified diagram of the nuclear states in $^{40}$Ar, showing one example of NC excitation by neutrinos and de-excitation by single gamma-ray emission.}
\label{fig:actionDiagram}
\end{figure}

Figure~\ref{fig:actionDiagram} sketches the nuclear physics of the NC interaction channel we consider,
\begin{equation}
\nu + \, ^{40}{\rm Ar} \rightarrow \nu + \, ^{40}{\rm Ar}^*,
\label{eq:argonNC}
\end{equation}
where * again indicates a nuclear excited state (we ignore subdominant final states where nucleons are ejected).  We consider only the most detectable final states, which have excitation energies of 5--10~MeV and decay directly to the ground state by the emission of \textit{single} gamma rays.  Because of the high nuclear thresholds, the detection rate is sensitive to only the high-energy tail of the $^8$B spectrum, where the electron neutrino survival probability is nearly energy-independent.  NC interactions are the same for all three neutrino flavors; they are somewhat different for neutrinos and antineutrinos, which could become important for exotic models of flavor change.

The total NC neutrino-nucleus cross section can be calculated from electroweak theory~\cite{PhysRevC.6.719}, which requires the nuclear matrix elements for the transitions of interest.  These can be calculated from first principles, using the nuclear Hamiltonian and many-body theory.  However, various theoretical approaches --- for example, the shell model~\cite{Dutta:2022tav} or the continuum random phase approximation~\cite{VanDessel:2020epd} --- show discrepancies between each other and data~\cite{Gayer:2019eed}. Notably, the shell model overestimates the magnetic dipole strength by a factor of $\sim$6~\cite{Tornow:2022kmo}. Therefore, we use a direct empirical approach where the nuclear matrix elements are extracted from data. This follows similar successful treatments of NC and CC cross sections for various nuclear targets for solar and supernova neutrinos, e.g., in Refs.~\cite{Bahcall:1987jc, Trinder:1997xr, Langanke:2004vx, Capozzi:2018ubv}.  While more detailed studies of the cross section will be needed to measure the absolute event rates, the formalism used here is adequate to establish the scale of the expected statistical uncertainty.

In this empirical approach, one considers the long-wavelength ($q\rightarrow 0$) limit~\cite{Langanke:2004vx,Dutta:2022tav} and only the final states that dominate the reaction, which are picked out based on angular momentum selection rules~\cite{Kolbe:2003ys}. Given that the ground state of ${^{40}}$Ar is $J^P=0^+$, the reaction is dominated by transitions to $J^P=1^+$ excited states; these are all allowed Gamow-Teller (GT) transitions.  In these limits, the cross section for the transition between the ground state and the $j^{\mathrm{th}}$ nuclear excited state is
\begin{equation}
    \sigma_{j} (E_\nu) = \frac{G_F^2}{\pi} (E_\nu - \omega_j)^2 B_j(GT).
    \label{eq:GT}
\end{equation}
Here $E_\nu$ is the neutrino energy, $\omega_j$ is the energy difference between the ground state and state $j$, $G_F$ is the Fermi constant, and $B_j(GT)$ is the reduced transition probability or strength. The axial-vector constant $g_A$ and other factors are absorbed into $B_j(GT)$~\cite{Tornow:2022kmo}.  This approach has been shown to accurately reproduce the total all-orders cross section from shell model calculations (after rescaling it to reproduce the correct $B(M1)$ strengths) in the energy range of solar neutrinos~\cite{Tornow:2022kmo}. 

Computing the cross section now amounts to determining the strengths, $B_j(GT)$.  This is achieved through measurements of the magnetic dipole strength, $B(M1)$, which, in the long-wavelength limit, is simply a scaling of $B_j(GT)$~\cite{Langanke:2004vx}. This approach gives: (i) an excellent measurement of the excited-state energy, (ii) a value for the strength with well-controlled uncertainties, and (iii) a prescription for how the excited state decays.

\begin{figure}[t]
\centering
\includegraphics[width=\columnwidth]{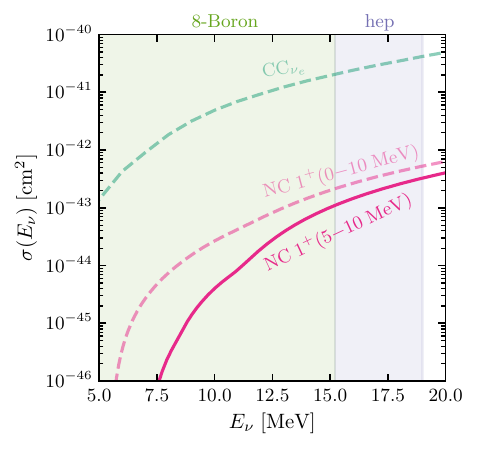}
\caption{NC neutrino-argon cross section for the 5--10 MeV $1^+$ excitations we consider.  For comparison, we also show the NC cross section including lower-energy $1^+$ excitations as well as the CC cross section.  We note the energy ranges where different solar fluxes dominate in the parent neutrino spectra.}.
\label{fig:xsecs}
\end{figure}

A first recent key development is new measurements of the $1^+$ states in $^{40}$Ar up to 10 MeV using nuclear resonance fluorescence~\cite{PhysRevC.73.054306, Gayer:2019eed, Tornow:2022kmo}.  These measurements of the $B(M1)$ strength found new states and reduced the total uncertainty on transitions to these states from 40\% to within 20\%, which could be improved with new experiments. These measurements also established that these states dominantly decay directly to the ground state through the emission of single gamma rays.  For the expected energy resolution of DUNE, some of these states are degenerate, though we calculate them separately. While higher-energy states may exist, they would only contribute minimally, due to the steeply falling solar-neutrino spectrum.

Figure~\ref{fig:xsecs} shows the NC cross section for 5--10 MeV $1^+$ excitations compared to the NC cross section including lower-energy $1^+$ excitations that lead to lower-energy gamma rays.  We also compare to the total CC cross section (for transitions to bound excited states in $^{40}$K) from Refs.~\cite{Gardiner:2020ulp, Gardiner:2021qfr} (which build on Refs.~\cite{Bhattacharya:2009zz, cheoun_high-lying_2012}).  The total CC cross section is much larger than the NC cross sections because the total strength for the CC nuclear transitions up to 15 MeV is $B(F+GT)=9.7$~\cite{PhysRevC.58.3677}, much larger than for the NC cases.

A second recent key development is that the prospects for detecting nuclear de-excitation gamma rays in liquid-argon detectors --- conservatively neglected in Ref.~\cite{Capozzi:2018dat} --- have recently improved.  In the energy range we consider, gamma rays produce detectable signals primarily through Compton scattering and pair production. ArgoNeuT has shown experimentally that they can reconstruct MeV-range nuclear de-excitation gamma rays in liquid argon~\cite{ArgoNeuT:2018tvi}. Further, Refs.~\cite{Castiglioni:2020tsu, Gardiner:2020ulp} show that these gamma rays should also be detectable in DUNE.


\section{Prospects for DUNE}
\label{sec:DUNE_prospects}

DUNE will leverage the transformative power of its Liquid Argon Time Projection Chamber (LArTPC) to reconstruct events with sub-cm-scale position resolution, allowing detailed tracking of individual particles~\cite{DUNE:2020lwj, DUNE:2020ypp, DUNE:2020mra, DUNE:2020txw}.  Because this position resolution is much smaller than the radiation length (about 14~cm in liquid argon), this makes it possible to distinguish different event classes.  For example, this is critical to DUNE's ability to separate NC and CC neutrino-argon events (as sketched in Fig.~\ref{fig:event_topology}).  DUNE will initially have two modules, with two more planned to be added later, with each module having a fiducial mass of 10 kton.  They will be located 1.5~km underground in the Sanford Underground Research Facility (SURF) in South Dakota.

To predict the number of NC events at DUNE, we convolve the arriving neutrino flux with the NC cross section.  We use the unoscillated (because this is a NC interaction) neutrino fluxes from the BS05(OP) solar model~\cite{Bahcall:1997eg, Winter:2004kf, Bahcall:2004pz, bahcallOnline}.  For a given Gamow-Teller transition ($j$), the number of NC events is
\begin{equation}
    \mathcal{N}^j_{NC} = N_{t} \, \Delta t \int_{\omega_{j}}^{\infty} \mathrm{d}E_\nu \, 
    \frac{\mathrm{d}\Phi_\mathrm{solar}}{\mathrm{d}E_\nu}(E_\nu) \, \sigma_{j}(E_\nu),
\label{eq:yield}
\end{equation}
where $N_t$ is the number of target argon atoms and $\Delta t$ is the detector livetime.  For the exposure, we assume 100~kton-year, equivalent to five years of livetime for the initial two-module configuration.  

The nuclear excitations that we consider decay predominantly via isotropic emission of a single high-energy gamma ray, corresponding to the nuclear energy level.  For a particular excitation, the yield is the only observable, i.e., the only information accessible is the \textit{integral} in Eq.~(\ref{eq:yield}).  This means that only the $^8$B flux matters, as it is much larger than the $hep$ flux (and the atmospheric-neutrino flux).  Information about the \textit{integrand} can be obtained by comparing the NC yields for different nuclear excitations or, more directly, via CC interactions.

\begin{figure}[t]
\centering
\includegraphics[width=\columnwidth]{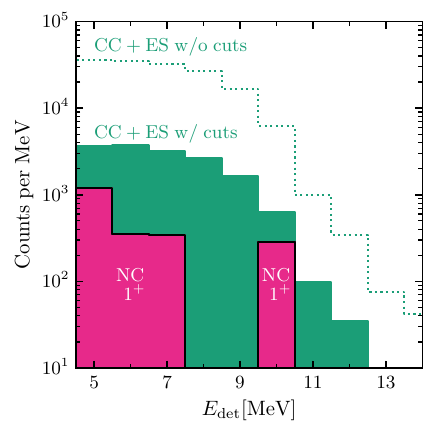}
\caption{Expected spectra for NC (fuschia) and CC+ES events (green) for a DUNE exposure of 100-kton-year. The solid regions show our predictions after PID and an angular cut, while the dotted line shows the CC+ES spectrum before cuts.}
\label{fig:counts}
\end{figure}

\begin{figure*}[t]
\centering
\includegraphics[width=\textwidth]{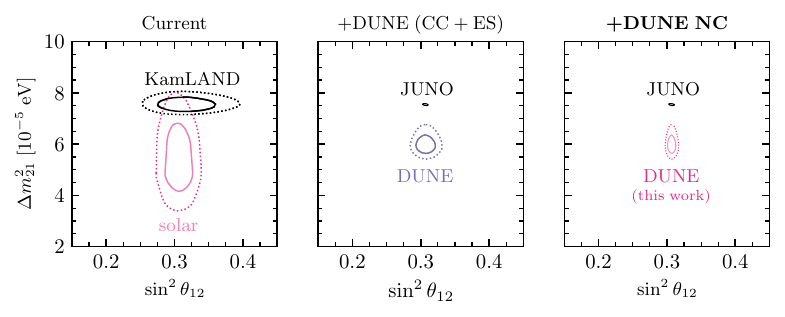}
\caption{Uncertainty contours (1- and 2-sigma) for the solar mixing parameters.  \textbf{Left panel:} Current results using the combined solar neutrino data (depending mostly on Super-Kamiokande and SNO)~\cite{deSalas:2020pgw} and the reactor antineutrino data of KamLAND~\cite{KamLAND:2010fvi}.  \textbf{Middle panel:} Predictions for DUNE using CC+ES events~\cite{Capozzi:2018dat} and for the JUNO reactor data~\cite{Capozzi:2015bpa}.  \textbf{Right panel:} \textit{Our predictions for DUNE based on using the solar NC events in addition to the CC+ES events.} The middle and right panels assume eventual cross-section uncertainties of 1\%.}
\label{fig:results}
\end{figure*}

Reducing backgrounds to the NC signal in DUNE will require dedicated but realistic work by experimentalists, as detailed in Refs.~\cite{Capozzi:2018dat, Parsa:2022mnj, Caratelli:2022llt}.  We focus our attention on backgrounds that are present in the fiducial volume, and thus not shielded by the outer layers of liquid argon.  First, the experimentalists must reduce the large background from the radiative capture of thermal neutrons on argon (releasing 6.1~MeV in $\sim$3 gamma rays~\cite{Lycklama:1967tex}).  This can be accomplished with at least 50~cm of hydrogenous shielding outside the detector~\cite{Capozzi:2018dat, Manzanillas:2024}; this shielding need not be perfectly hermetic because of the random motions of neutrons.  Second, they must reduce the large background from the $(\alpha,\gamma)$ process on argon, where the $\alpha$ particles arise from dissolved $^{222}$Rn and where the gamma-ray energies are up to 15~MeV~\cite{Reichenbacher2018_talk, Caratelli:2022llt, Westerdale2024_talk}.  This can be accomplished through standard radon-mitigation techniques, which have reduced the concentrations in, e.g., Super-Kamiokande, SNO, and MicroBooNE by several orders of magnitude~\cite{SuperKamiokade:1999bcy, Blevis:2003ih, MicroBooNE:2022his, MicroBooNE:2023sxs}.  Third, they must reduce the spallation background to the estimates of Ref.~\cite{Zhu:2018rwc} by employing the ideas noted there.

With the steps above, the dominant background would be that induced by solar-neutrino CC+ES interactions.  For this, we follow Ref.~\cite{Capozzi:2018dat}, including taking into account neutrino mixing in the Sun and Earth, which gives an energy-independent suppression during the day and an energy-dependent change during the night.  The CC interactions predominantly result in an electron and a $^{40}$K nucleus in a nuclear excited state, with a subleading ($< 1$\%) component with a final-state neutron and $^{39}$K~\cite{Gardiner:2020ulp}.  The excited states have energies up to 8~MeV and de-excite via the isotropic emission of \textit{multiple} gamma rays whose energies range between 30~keV and 2~MeV~\cite{2011PhRvC..83b8801C, nuclearData}.  These low energies make the CC de-excitation gamma rays harder to detect, as they produce ``blips" within the detector, corresponding to low-energy Compton-scattered electrons~\cite{Castiglioni:2020tsu, Gardiner:2020ulp}. This is in contrast to higher energy \textit{single} $^{40}$Ar de-excitation gamma rays from NC interactions, which are easier to detect because they produce longer electron tracks. The ES interactions result in just a single electron, which is scattered forward.

To suppress CC+ES backgrounds, we first use particle identification (PID) techniques developed to distinguish gamma rays from electrons via their energy-loss rates~\cite{ArgoNeuT:2016wjb, DUNE:2020ypp, Castiglioni:2020tsu, Gardiner:2020ulp, Newmark:2023vup}.  Next, we apply an angular cut to remove most of the ES background~\cite{Capozzi:2018dat}.  In combination, these two cuts retain 80\% of the signal but only 10\% of the background.  With dedicated studies --- including using topological classifications~\cite{Castiglioni:2020tsu, Gardiner:2020ulp} and tagging specific long-lived state of $^{40}$K~\cite{nuclearData} --- we expect that the cuts could be much improved. Details concerning PID and the cuts are given in Appendix~\ref{app:cuts}. With this approach, the signal would be smaller than the background but larger than its uncertainty.  Further background reduction would be helpful but not essential.

Figure~\ref{fig:counts} shows the expected spectra in DUNE for an exposure of 100~kton-year, where we require detected energies to be above 4.5~MeV.  For the $x$-axis, we follow Ref.~\cite{Anderson:2012vc} and define the detectable energy as $E_\mathrm{det} = E_e + \Sigma E_\gamma$, where only tracks that have an energy deposit above 300~keV at a single point are included, though DUNE may be able to do better~\cite{Castiglioni:2020tsu, Gardiner:2020ulp}.  (In contrast, Ref.~\cite{Capozzi:2018dat} conservatively assumed that gamma rays were undetectable.)  \textit{After cuts}, we predict $\simeq$2200 NC events in total above 4.5~MeV and $\simeq$15000 CC+ES events above 4.5~MeV. The small peak at 6 MeV in the CC+ES spectrum after cuts is caused by $\simeq$240 captured neutrons from the rare branch of CC events with a final-state neutron.

To estimate DUNE's sensitivity to the solar mixing parameters, we take both statistical and systematic uncertainties into account and perform a $\chi^2$ fit for $\Phi_\mathrm{solar}$ and $\sin^2\theta_{12}$, because $\Delta m^2_{21}$ will be determined independently via the day-night effect in the CC+ES data~\cite{Capozzi:2018dat}.  Without PID and the angular cut, the dataset is dominated by CC+ES events, which allow a high-precision measurement of the product $\Phi_\mathrm{solar} \sin^2\theta_{12}$. With PID and the angular cut, the dataset has an enhanced NC fraction that allows us to break the degeneracy between those two factors.

For the fits, we generate $10^6$ simulated datasets, each corresponding to an exposure of 100~kton-year, including the additional backgrounds discussed above.  We assume a standard $^8$B spectrum shape~\cite{Ortiz:2000nf}, with 2\% uncertainty on the normalization of the $\nu_e$ flux~\cite{Super-Kamiokande:2023jbt} and include an energy reconstruction uncertainty of $\sigma / E \sim 10\% / \sqrt{E\mathrm{[MeV]}}\oplus 2\%$~\cite{LArIAT:2019gdz, Castiglioni:2020tsu}.  (Our results are not very sensitive to the assumed energy, angular, or position resolutions of DUNE.)  We then perform a $\chi^2$ fit on the datasets without and with PID and the angular cut.  We use the resulting distributions of the best-fit values to calculate the mean and confidence intervals of the fit parameters. The final uncertainty is statistics-dominated due to the relatively low counts of the NC events.  We assume that the neutrino-argon NC cross sections could eventually be measured to 1\% precision, which is ambitious but not out of reach, given the extremely high scientific payoffs (in Appendix~\ref{app:systematics}, we show results assuming a larger cross-section uncertainty.)  An eventual precision of 1\% is the same that Ref.~\cite{Capozzi:2018dat} assumed for the neutrino-argon CC cross section and for which they discussed a detailed program of experimental and theoretical work to reach it.  Importantly, the NC and CC cross sections can be determined through precise measurements of nuclear transitions as well as with laboratory neutrino experiments.

Figure~\ref{fig:results} shows how the precision of the solar mixing parameters improves at successive steps, starting with the current allowed regions, then with the predictions of Ref.~\cite{Capozzi:2018dat}, and last with our results.  For the central values, we use the current best fits of Super-Kamiokande \cite{Super-Kamiokande:2023jbt} and KamLAND~\cite{KamLAND:2010fvi}.  \textit{The primary benefit of including the NC events is that this improves the uncertainty on $\sin^2\theta_{12}$ from $4.5\%$ to 1.7\%.} Additionally, it improves the uncertainty on the total solar neutrino flux from 5\%~\cite{SNO:2011hxd} to 1.6\%.  For $\Delta m^2_{21}$, the day-night effect used in Ref.~\cite{Capozzi:2018dat} remains the best approach.  Importantly, using CC, ES, and NC measurements in DUNE would make its solar-neutrino uncertainties much more comparable to JUNO's reactor-antineutrino uncertainties, allowing unprecedented tests of new physics.  At present, there is a slight tension between the solar and reactor $\Delta m^2_{21}$ values.  Visually, this makes the $\Delta m^2_{21}$ comparison more noticeable, but it is important to compare precision measurements of \textit{both} $\sin^2\theta_{12}$ and $\Delta m^2_{21}$.

We restricted our analysis to excitations in the range 5--10 MeV. Extending this range would improve our results.  Known transitions below 4.5 MeV~\cite{Tornow:2022kmo} would enable the NC dataset to be approximately doubled, plus theory predicts another strong transition at $11\;\mathrm{MeV}$.  Additionally, since the projected uncertainty is statistics-limited, additional modules would help.


\section{Conclusions and Outlook}
\label{sec:conclusion}

We have shown that DUNE can measure the total active-flavor flux of $^8$B solar neutrinos using NC interactions with argon. Because DUNE can also measure the surviving $\nu_e$ flux using CC interactions with argon, they would thus precisely measure $\sin\theta_{12}$.  The possibility of improving upon SNO's NC and CC solar-neutrino measurements is unique to DUNE.  Further, through the day-night effect in the CC data, DUNE would also improve the precision of $\Delta m^2_{21}$ over all previous solar measurements.  Comparison of the neutrino mixing parameters measured with DUNE and JUNO would allow unprecedented tests of new physics in solar neutrinos~\cite{Barbier:2004ez, Friedland:2004pp, Palazzo:2011rj, Barenboim:2023krl, deGouvea:2023jxn, Ansarifard:2024zxm, Wu:2023twu, Denton:2024upc}.

The precise measurement of the total $^8$B flux would help probe the temperature and metallicity of the solar core~\cite{Bahcall:1996vj, Wurm:2017cmm}. This could help resolve a longstanding discrepancy between two different methods of determining the solar metallicity~\cite{Basu:2007fp, Bergemann:2014vaa, Christensen-Dalsgaard:2020imv, Gann:2021ndb}.

Realizing DUNE's potential for these solar-neutrino measurements is challenging but realistic. It will require investments in background reduction~\cite{Zhu:2018rwc, Capozzi:2018dat, Gardiner:2020ulp, Manzanillas:2024}, argon cross section measurements~\cite{Grant:2015jva, Baxter:2019mcx, Barbeau:2021exu, CCM:2021leg, JSNS2:2021hyk, Kelly:2021jgj, DUNE:2023rtr}, and event-selection studies.  The costs of these are very likely far below the cost of DUNE, but would substantially improve its scientific impact.


\section*{Acknowledgements}

We are grateful for helpful discussions with Francesco Capozzi, David Caratelli, Shirley Li, Vishvas Pandey, and Austin Schneider.  This work was supported by the Australian Research Council through Discovery Project DP220101727 plus the University of Melbourne’s Research Computing Services and the Petascale Campus Initiative. J.L.N was supported by the ARC Centre of Excellence for Dark Matter Particle Physics, CE200100008.  J.F.B. was supported by US National Science Foundation Grant No.\ PHY-2310018.  J.F.B. speaks for himself as a theorist, not on behalf of the DUNE Collaboration. This work is based on the ideas and calculations of the authors, plus publicly available information.


\appendix
\counterwithin{figure}{section}
\vspace{0.75cm}
\centerline{\Large {\bf Appendices}}
\vspace{0.75cm}

Here, we provide additional details that may be helpful. We discuss particle-identification techniques (Appendix~\ref{app:cuts}) and how our results depend upon the NC cross-section uncertainty (Appendix~\ref{app:systematics}).h

\section{Particle Identification}
\label{app:cuts}

We implement a basic form of particle identification (PID) to suppress backgrounds.  Using {\tt GEANT4}~\cite{GEANT4:2002zbu}, we track the propagation of final-state particles from NC, CC, and ES interactions.  The key task is to distinguish gamma rays from electrons, taking of the excellent position resolution (sub-cm) of liquid-argon detectors.

There are two relevant cases.  First, the gamma ray deposits all of its energy via Compton scattering, producing multiple low-energy electrons, which typically appear as multiple discrete points or ``blips"~\cite{Castiglioni:2020tsu, Gardiner:2020ulp} in the detector.  Second, when it converts to an electron-positron pair, which typically leads to higher-energy electrons that produce recognizable tracks.  The likelihood of a gamma ray converting to an electron-positron pair rises from $<30\%$ at 5~MeV to $>80\%$ at 10~MeV~\cite{PDG2024}.  We follow Refs.~\cite{ArgoNeuT:2016wjb, DUNE:2020ypp, Newmark:2023vup}, who showed that ionizing electron-positron pairs could be cleanly separated from single ionizing electrons by their $|dE/dx|$ values.  The differences in $|dE/dx|$ arise from the number of minimally ionizing particles, each having an expected energy loss of 2~MeV/cm.  We find that cutting at 3.1~MeV/cm retains 95\% of the NC events and only 16\% of the CC+ES events.  When a gamma ray does not produce a pair, we assume a 100\% separation efficiency. 


\section{Systematics}
\label{app:systematics}

\begin{figure*}[t]
\centering
\includegraphics[width=\textwidth]{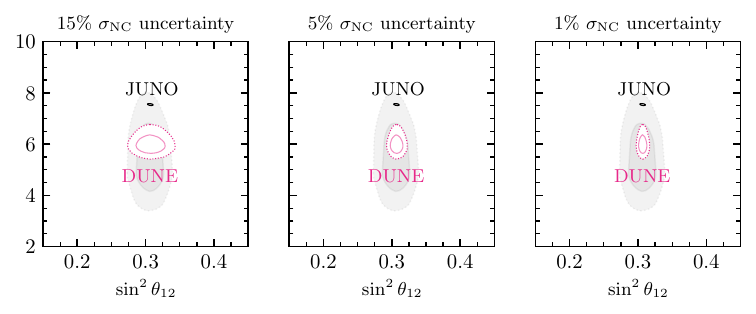}
\caption{Same as Figure~\ref{fig:results} right, with 15\%, 5\% and 1\% uncertainties on the NC neutrino-argon cross section (and an assumed 1\% uncertainty on the CC neutrino-argon cross section, following Ref.~\cite{Capozzi:2018dat}).  In faint gray, we show the contours for the present solar mixing parameters.}
\label{fig:results_sys}
\end{figure*}

Figure~\ref{fig:results_sys} shows how the precision on $\sin^2\theta_{12}$ depends on the assumed uncertainty on the NC neutrino-argon cross section.  We show three cases, assuming 15\% (close to the current uncertainty~\cite{Gayer:2019eed, Tornow:2022kmo}), 5\% (a reasonable improvement), and 1\% (the ideal case).  Once the cross section uncertainty is below $\sim$3\%, it is no longer a limiting factor compared to event statistics.

When aiming for 1\% precision, other factors will become important, including the uncertainty on the size of the fiducial volume, which can be largely overcome by considering the ratios of the CC, ES, and NC rates.

\clearpage
\bibliography{bibliography}


\end{document}